\documentclass[journal]{IEEEtran}
\IEEEoverridecommandlockouts
\usepackage{lineno}
\usepackage{hyperref}
\usepackage{cite}
\usepackage{amsmath,amssymb,amsfonts,cases} 
\usepackage{amsmath}
\usepackage{amsthm}
\DeclareMathOperator*{\argmax}{argmax}
\DeclareMathOperator*{\argmin}{argmin}
\usepackage{graphicx}
\usepackage{textcomp}
\usepackage{xcolor}
\usepackage{graphicx}
\usepackage{float}
\usepackage{subfigure}
\usepackage{amsmath,mleftright}
\usepackage{amsfonts,amssymb}
\usepackage{mathrsfs}
\usepackage{mathtools}
\usepackage{algorithm}
\usepackage{algorithmic}
\usepackage{bm}
\usepackage{multirow}
\usepackage{array}
\usepackage{amssymb}
\usepackage{amsmath}
\usepackage{cite}
\usepackage{url}
\usepackage{xcolor}
\usepackage{cite,graphicx,amsmath,amssymb}
\usepackage{subfigure}
\usepackage{fancyhdr}
\usepackage{mdwmath}
\usepackage{mdwtab}
\usepackage{caption}
\usepackage{amsthm}
\usepackage{setspace}
\usepackage{bm}
\usepackage{mathtools}
\usepackage{dsfont}
\usepackage{bbm}
\usepackage{framed}

\newtheorem{theorem}{Theorem}

\newtheorem{lemma}{Lemma}

\newtheorem{corollary}{Corollary}

\makeatletter
\newcommand{\biggg}{\bBigg@{3}}
\newcommand{\Biggg}{\bBigg@{3.5}}
\makeatother
\makeatletter
\allowdisplaybreaks[4]
\renewcommand{\maketag@@@}[1]{\hbox{\m@th\normalsize\normalfont#1}}%
\makeatother
\def\BibTeX{{\rm B\kern-.05em{\sc i\kern-.025em b}\kern-.08em
    T\kern-.1667em\lower.7ex\hbox{E}\kern-.125emX}}
    \expandafter\def\expandafter\normalsize\expandafter{%
    \normalsize%
    \setlength\abovedisplayskip{4pt}%
    \setlength\belowdisplayskip{4pt}%
    \setlength\abovedisplayshortskip{2pt}%
    \setlength\belowdisplayshortskip{2pt}%
}
\begin{document}
\title{Sum-Rate Maximization for Uplink Segmented Waveguide-Enabled Pinching-Antenna Systems}
\author{Songnan~Gu, Hao~Jiang, Chongjun~Ouyang, Yuanwei~Liu, and Dong~In~Kim\vspace{-10pt}
\thanks{S. Gu, H. Jiang, and C. Ouyang are with the School of Electronic Engineering and Computer Science, Queen Mary University of London, London, E1 4NS, U.K. (e-mail: s.gu@se23.qmul.ac.uk and \{hao.jiang, c.ouyang\}@qmul.ac.uk).}
\thanks{Y. Liu is with the Department of Electrical and Electronic Engineering, The University of Hong Kong, Hong Kong (email: yuanwei@hku.hk).}
\thanks{D. I. Kim is with the Department of Electrical and Computer Engineering, Sungkyunkwan University, Suwon 16419, South Korea (e-mail: dongin@skku.edu).}}
\maketitle
\begin{abstract}
A multiuser uplink transmission framework based on the segmented waveguide-enabled pinching-antenna system (SWAN) is proposed under two operating protocols: segment selection (SS) and segment aggregation (SA). For each protocol, the achievable uplink sum-rate is characterized for both time-division multiple access (TDMA) and non-orthogonal multiple access (NOMA). Low-complexity placement methods for the pinching antennas (PAs) are developed for both protocols and for both multiple-access schemes. Numerical results validate the effectiveness of the proposed methods and show that SWAN achieves higher sum-rate performance than conventional pinching-antenna systems, while SA provides additional performance gains over SS.
\end{abstract}
\section{Introduction}
With the rapid development and commercialization of fifth-generation (5G) networks, the forthcoming sixth-generation (6G) systems are expected to surpass 5G across all major performance dimensions. To meet these expectations, reconfigurable-antenna technologies, including fluid and movable antennas \cite{you2025next}, have attracted considerable attention because they can reshape the wireless channel and establish more favorable propagation conditions. However, conventional reconfigurable-antenna systems typically support reconfiguration only on the wavelength scale, which limits their ability to overcome large-scale path loss and signal blockage \cite{liu2025pinching}. Moreover, repositioning antennas in these systems often incurs high implementation cost and difficulty \cite{liu2025pinching}.

To address these limitations, NTT DOCOMO introduced the \emph{pinching-antenna system (PASS)} at the Mobile World Congress in 2021 \cite{suzuki2022pinching}. PASS consists of two main components: \emph{dielectric waveguides} and detachable dielectric \emph{pinching antennas (PAs)}. The waveguides serve as transmission media that deliver signals over long distances, often on the order of tens of meters, with low attenuation \cite{suzuki2022pinching,ding2024flexible}. By attaching PAs at selected positions along the waveguides, the guided signals can radiate into or from free space at desired locations \cite{suzuki2022pinching,ding2024flexible}. In contrast to other reconfigurable-antenna technologies that offer only wavelength-scale flexibility, PASS provides meter-scale reconfigurability through the extended physical reach of the waveguides. This enables flexible PA activation at many candidate positions and allows PASS to mitigate both large-scale path loss and signal blockage. In essence, PASS can ``pass'' signals directly to users by placing active PAs near their locations. This distinctive capability has drawn significant research interest \cite{liu2025pinching1}.

\begin{figure}[!t]
\centering
    \subfigure[Segmented waveguide.]
    {
        \includegraphics[width=0.35\textwidth]{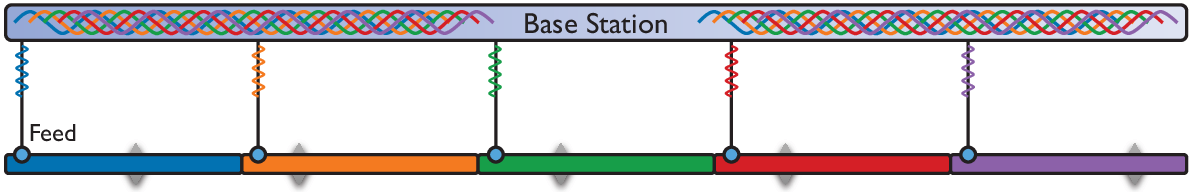}
	   \label{Figure: PAS_System_Model2}
    }
    \subfigure[System setup.]
    {
        \includegraphics[width=0.35\textwidth]{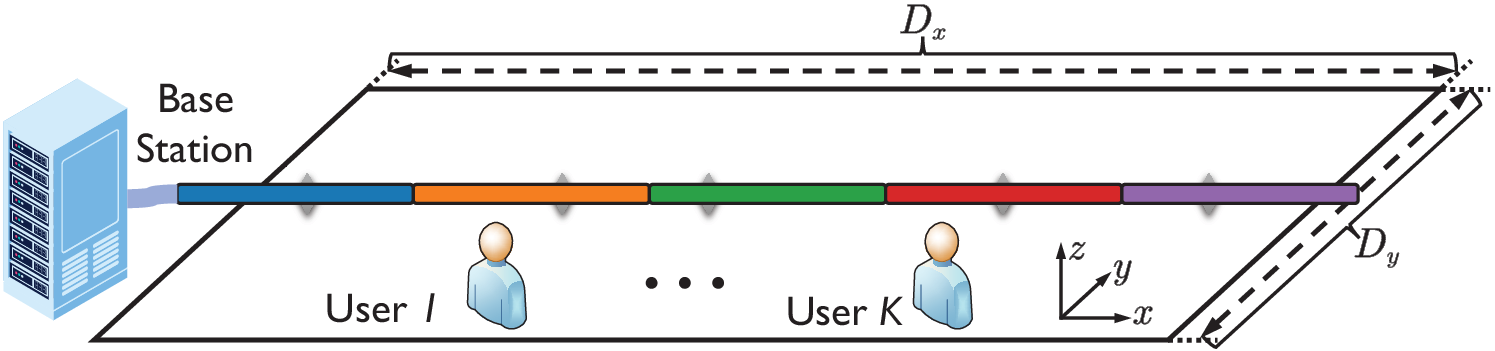}
	   \label{Figure: PAS_System_Model1}
    }
\caption{Illustration of the SWAN-based uplink channel.}
\label{Figure: PAS_System_Model}
\vspace{-15pt}
\end{figure}

Despite the growing body of research on PASS \cite{liu2025pinching1}, a key challenge remains in the current PASS architecture, namely the lack of a tractable uplink model for multiple active PAs. When multiple PAs are deployed along the same waveguide, the signal captured by one PA can re-radiate through other PAs as it propagates toward the feed point. This \emph{inter-antenna radiation (IAR)} effect complicates uplink analysis and leads to mathematically intractable signal models. As a result, most studies on uplink PASS either restrict attention to single-PA deployments, which naturally avoid IAR, or ignore the IAR effect entirely when multiple PAs are present \cite{liu2025pinching1}. A recent effort to resolve this issue introduced the \emph{segmented waveguide-enabled pinching-antenna system (SWAN)} in \cite{ouyang2025uplinkSWAN}. Unlike conventional PASS that uses a single long waveguide, SWAN employs a set of short waveguide segments that are placed end-to-end but are \emph{physically unconnected}. Each segment features its own feed point, and each feed point connects to the base station (BS) through a dedicated wired link such as an optical fiber, as shown in {\figurename} {\ref{Figure: PAS_System_Model2}}. By activating one PA per segment, SWAN eliminates IAR and produces a tractable uplink multiple-PA model \cite{ouyang2025uplinkSWAN}.

The study in \cite{ouyang2025uplinkSWAN} focused solely on the single-user case. The design of transmission protocols and the associated PA placement strategies for multiuser uplink SWAN systems remains an open and unexplored problem. To address this research gap, this article investigates multiuser uplink communications in SWAN. We establish a multiuser SWAN transmission framework that incorporates two operating protocols: \emph{segment selection (SS)} and \emph{segment aggregation (SA)}. Both protocols are examined under time-division multiple access (TDMA) and non-orthogonal multiple access (NOMA). For SS-based transmission, we develop optimal or near-optimal PA placement methods that maximize the uplink sum-rate. For SA-based transmission, we introduce several low-complexity PA placement algorithms based on element-wise alternating optimization to enhance the sum-rate performance. Numerical results demonstrate that both SS-based and SA-based SWAN outperform conventional uplink PASS in terms of achievable sum-rate under both TDMA and NOMA.
\section{System Model}
Consider an uplink channel where a BS employs a segmented waveguide to serve $K$ users located in a rectangular service region with dimensions $D_x$ and $D_y$ along the $x$- and $y$-axes. The position of the $k$th user is denoted by ${\mathbf{u}}_k = [u_k^x, u_k^y, 0]^{\mathsf{T}}$ for $k\in{\mathcal{K}}\triangleq\{1,\ldots,K\}$. The waveguide is deployed along the $x$-axis to provide horizontal coverage, with PAs placed along it to receive signals; see {\figurename} {\ref{Figure: PAS_System_Model1}}.

The segmented waveguide contains $M$ dielectric waveguide segments. Each segment has length $L$ such that $D_x = M L$. Let ${\bm\psi}_{0}^{m}\triangleq[\psi_{0}^{m},0,d]^{\mathsf{T}}$ denote the location of the feed point of the $m$th segment for $m\in{\mathcal{M}}\triangleq\{1,\ldots,M\}$, where $\psi_{0}^{1}<\psi_{0}^{2}<\ldots<\psi_{0}^{M}$, and $d$ represents the deployment height of the waveguide. For simplicity, the feed point is placed at the front-left end of each segment. To avoid uplink IAR, only one PA is activated in each segment. The position of the PA in the $m$th segment is denoted by ${\bm\psi}_{m}\triangleq[\psi_{m},0,d]^{\mathsf{T}}$, and satisfies 
\begin{align}
\psi_{0}^{m}\leq \psi_{m}\leq \psi_{0}^{m}+L,\lvert\psi_{m}-\psi_{m'}\rvert\geq \Delta,\forall m\ne m',
\end{align}
where $\Delta>0$ is the minimum inter-antenna spacing required to mitigate mutual coupling effects.

\subsection{Channel Model}
PASS is envisioned for operation in high-frequency bands \cite{suzuki2022pinching}, where line-of-sight (LoS) propagation dominates \cite{ouyang2024primer}. A free-space LoS channel model is therefore adopted. Under this model, the spatial channel coefficient between the $m$th PA and user $k$ can be characterized as follows \cite{ouyang2024primer}:
\begin{align}
h_{\rm{o}}({\mathbf{u}}_k,{\bm\psi}_{m})\triangleq
\frac{\eta^{\frac{1}{2}}{\rm{e}}^{-{\rm{j}}k_0\lVert{\mathbf{u}}_k-{\bm\psi}_{m}\rVert}}{\lVert{\mathbf{u}}_k-{\bm\psi}_{m}\rVert},
\end{align}
where $\eta\triangleq\frac{c^2}{16\pi^2f_{\rm{c}}^2}$, $c$ is the speed of light, $f_{\rm{c}}$ is the carrier frequency, $\lambda$ is the free-space wavelength, and $k_0=\frac{2\pi}{\lambda}$ is the wavenumber. Furthermore, the in-waveguide propagation coefficient between the feed point and the $m$th PA follows the model in \cite{pozar2021microwave}:
\begin{align}\label{In_Waveguide_Channel_Model}
h_{\rm{i}}({\bm\psi}_{m},{\bm\psi}_{0}^{m})&\triangleq{10^{-\frac{\kappa}{20}\lVert{\bm\psi}_{m}-{\bm\psi}_{0}^{m}\rVert}}
{\rm{e}}^{-{\rm{j}}\frac{2\pi\lVert{\bm\psi}_{m}-{\bm\psi}_{0}^{m}\rVert}{\lambda_{\rm{g}}}},
\end{align}
where $\lambda_{\rm{g}}=\frac{\lambda}{n_{\rm{eff}}}$ is the guided wavelength and $n_{\rm{eff}}$ is the effective refractive index of the dielectric waveguide \cite{pozar2021microwave}. The parameter $\kappa$ denotes the average attenuation factor along the waveguide in dB/m \cite{yeh2008essence}. The special case $\kappa=0$ corresponds to a lossless dielectric. Prior studies have shown that in-waveguide attenuation has \emph{negligible impact} on the performance of SWAN \cite{ouyang2025uplinkSWAN}. Based on this conclusion, we set $\kappa=0$ in the design. Performance under nonzero attenuation is evaluated through numerical experiments in Section \ref{Section: Numerical Results}.

\begin{figure}[!t]
\centering
    \subfigure[Segment selection.]
    {
        \includegraphics[height=0.115\textwidth]{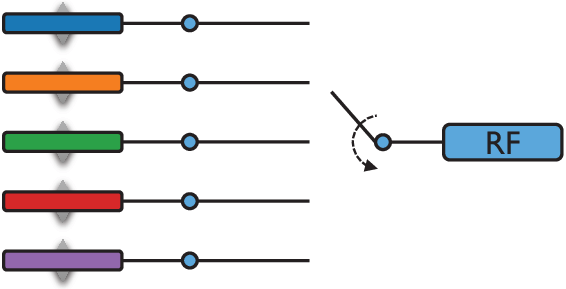}
	   \label{Figure_PAN_Protocol1}
    }
   \subfigure[Segment aggregation.]
    {
        \includegraphics[height=0.115\textwidth]{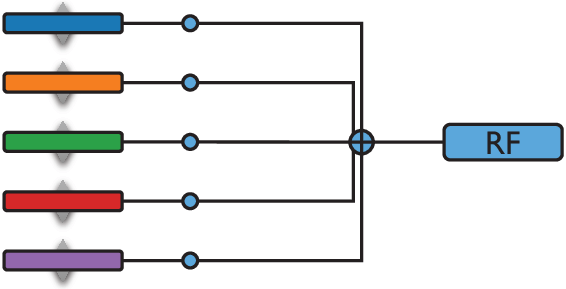}
	   \label{Figure_PAN_Protocol2}
    }
\caption{Illustration of the SS and SA protocols.}
\label{Figure: PAN_Protocol}
\vspace{-15pt}
\end{figure}

\subsection{Signal Model}
The performance of SWAN depends on how the feed points connect to the radio-frequency (RF) front-end of the BS, as shown in {\figurename} {\ref{Figure: PAS_System_Model2}}. This work considers two operating protocols: SS and SA, illustrated in {\figurename} {\ref{Figure: PAN_Protocol}}.
\subsubsection{Segment Selection}
In the SS protocol, only one segment is connected to the RF chain at any time, as shown in {\figurename} {\ref{Figure_PAN_Protocol1}}. Let ${\bar{m}}\in{\mathcal{M}}$ denote the index of the selected segment. The uplink signal received from user $k$ is $\sqrt{P_k}h_ks_k$, where $s_k\sim{\mathcal{CN}}(0,1)$ is the normalized data symbol and $P_k$ is the transmit power. The effective uplink channel of user $k$ is 
\begin{align}
h_k=h_{\rm{i}}({\bm\psi}_{{\bar{m}}},{\bm\psi}_{0}^{{\bar{m}}})h_{\rm{o}}({\mathbf{u}}_k,{\bm\psi}_{{\bar{m}}}),~\forall k\in{\mathcal{K}}.
\end{align}
Both TDMA and NOMA schemes are supported. In TDMA, the time axis is divided into orthogonal slots and each user transmits in a dedicated slot without interference. The corresponding sum-rate is given by
\begin{equation}\label{TDMA_Rate_Uplink}
    {\mathcal{R}}_{\rm{T}} = \frac{1}{K} \sum\nolimits_{k=1}^{K} \log_2 ( 1 + {P_k}\lvert h_k\rvert^2/{\sigma^2}  ),
\end{equation}
where $\sigma^2$ denotes the noise power at each PA. The channel gain of user $k$ can be written as follows: 
\begin{align}\label{Channel_Gain}
\lvert h_k\rvert^2=\frac{\eta}{\lVert{\mathbf{u}}_k-{\bm\psi}_{{\bar{m}}}\rVert^2}=\frac{\eta}{(u_k^x-{\psi}_{{\bar{m}}})^2+d_k},~\forall k\in{\mathcal{K}}, 
\end{align}
where $d_k\triangleq d^2+(u_k^y)^2$. Under NOMA with successive interference cancellation (SIC), all users transmit in the same time-frequency resources. The achievable sum-rate is given by
\begin{equation}\label{NOMA_Rate_Uplink}
    {\mathcal{R}}_{\rm{N}} = \log_2 \left( 1 + \sum\nolimits_{k=1}^{K} {P_k}\lvert h_k\rvert^2/{\sigma^2} \right).
\end{equation}
\subsubsection{Segment Aggregation}
The SS protocol relies on a simple switching mechanism and introduces very low hardware complexity. To exploit the array gain created by activating all segments simultaneously, this work also considers the SA protocol shown in {\figurename} {\ref{Figure_PAN_Protocol2}}. In SA, all feed points connect to a single RF chain through a signal combiner. The effective channel of user $k$ becomes
\begin{align}\label{MultiPAChannel}
h_k=\frac{1}{\sqrt{M}}\sum\nolimits_{m=1}^{M}h_{\rm{i}}({\bm\psi}_{{{m}}},{\bm\psi}_{0}^{{{m}}})h_{\rm{o}}({\mathbf{u}}_k,{\bm\psi}_{{{m}}}),~\forall k\in{\mathcal{K}},
\end{align}
where the factor $\frac{1}{\sqrt{M}}$ accounts for the cumulative noise from the $M$ segments. Under SA, the achievable sum-rates follow the same expressions as in \eqref{TDMA_Rate_Uplink} or \eqref{NOMA_Rate_Uplink}. 
\section{Optimization of the PA Placement}
This section develops efficient algorithms that optimize PA locations and maximize the achievable sum-rate for both the SS and SA protocols under TDMA and NOMA. For TDMA, two design schemes are considered. The first scheme optimizes the PA placement for each user within its scheduled slot. This scheme is referred to as \emph{pinch switching (PS)}. The second scheme requires all users to share a common PA placement across all slots, and is referred to as \emph{pinch multiplexing (PM)}. PS provides the maximum performance because the PA location can adapt to each individual user. PM requires a fixed PA configuration and thus reduces implementation and control complexity. PM also produces a conservative sum-rate and can be viewed as a practical lower-complexity alternative.
\subsection{Optimization for Segment Selection}
\subsubsection{PS-TDMA}
Under PS-based TDMA, the PA placement is optimized for each user within the user’s allocated time slot. With this design, sum-rate maximization is equivalent to maximizing the rate of each individual user in its slot. Referring to \eqref{TDMA_Rate_Uplink} and \eqref{Channel_Gain}, the corresponding problem for determining the PA location in the $k$th slot is formulated as follows: 
\begin{subequations}\label{PS_TDMA_SS}
\begin{align}
&\argmax\nolimits_{{\psi}_{m}\in[\psi_{0}^{m},\psi_{0}^{m}+L]}
\lvert h_{\rm{i}}({\bm\psi}_{{{m}}},{\bm\psi}_{0}^{{{m}}})h_{\rm{o}}({\mathbf{u}}_k,{\bm\psi}_{{{m}}})\rvert^2\\
&=\argmin\nolimits_{{\psi}_{m}\in[\psi_{0}^{m},\psi_{0}^{m}+L]}(u_k^x-{\psi}_{m})^2.
\end{align}
\end{subequations}
The optimal PA is the one activated in the segment closest to user $k$. The index of this segment is $m^{\star}=\lceil({u_k^x-{\psi}_{0}^{1}})/{L}\rceil$, and the optimal PA position aligns with the projection of the user onto the waveguide, i.e., ${\psi}_{m^{\star}}=u_k^x$.
\subsubsection{PM-TDMA}
Under PM-based TDMA, a single PA location is used for all time slots. The goal is to maximize the overall sum-rate, which is formulated as follows:
\begin{subequations}\label{PM_TDMA_SS}
\begin{align}
&\argmax_{{\psi}_{m}\in[\psi_{0}^{1},\psi_{0}^{1}+D_x]}
\sum\nolimits_{k=1}^{K} \log_2 \left( 1 + \frac{P_k\eta/{\sigma^2}}{(u_k^x-{\psi}_{m})^2+d_k}  \right)\\
&=\argmax_{{\psi}_{m}\in[\psi_{0}^{1},\psi_{0}^{1}+D_x]}
\prod\nolimits_{k=1}^{K} \left(1 + \frac{P_k\eta/{\sigma^2}}{(u_k^x-{\psi}_{m})^2+d_k}\right).
\end{align}
\end{subequations}
The problem reduces to maximizing a continuous single-variable function over a compact interval. The optimum appears either at an endpoint or at a point where the derivative equals zero. Newton's method can be applied to approximate such a stationary point. An alternative approach is a \emph{one-dimensional search}. The interval $[\psi_{0}^{1},\psi_{0}^{1}+D_x]$ is discretized into a uniform $Q$-point grid as follows:
\begin{align}
{\mathcal{Q}}\triangleq\left\{\psi_{0}^{1},\psi_{0}^{1}+\frac{D_x}{Q-1},\psi_{0}^{1}+\frac{2D_x}{Q-1},\ldots,\psi_{0}^{1}+D_x\right\}.
\end{align}
and the best point is selected by evaluating the objective at all grid positions. 
\subsubsection{NOMA}
Referring to \eqref{Channel_Gain} and \eqref{NOMA_Rate_Uplink}, the PA placement problem under SS and NOMA can be formulated as follows:
\begin{align}\label{NOMA_SS}
\argmax_{{\psi}_{m}\in[\psi_{0}^{1},\psi_{0}^{1}+D_x]}
\sum_{k=1}^{K} \frac{P_k}{(u_k^x-{\psi}_{m})^2+d_k}.
\end{align}
The solution can be approximated in the same manner as problem \eqref{PM_TDMA_SS}. Newton's method can be used to identify the stationary points, or a one-dimensional search can be performed over the interval $[\psi_{0}^{1},\psi_{0}^{1}+D_x]$. The computational complexity of the one-dimensional search for both TDMA and NOMA scales as ${\mathcal{O}}(QK)$.
\subsection{Optimization for Segment Aggregation}
We now optimize the PA placement under the SA protocol.
\subsubsection{PS-TDMA}
When PS is used under TDMA, the PA locations must be optimized for each slot to maximize the corresponding user's rate or signal-to-noise ratio (SNR). Referring to \eqref{MultiPAChannel}, the problem of optimizing the PA placement for the $k$th user can be formulated as follows:
\begin{subequations}\label{PS_TDMA_SA}
\begin{align}
&\max_{{\bm\psi}\in{\mathcal{X}}}
\left\lvert \sum\nolimits_{m=1}^{M}h_{\rm{i}}({\bm\psi}_{1}^{m},{\bm\psi}_{0}^{m})
h_{\rm{o}}({\mathbf{u}},{\bm\psi}_{1}^{m})\right\rvert^2\\
=&\max_{{\bm\psi}\in{\mathcal{X}}}
\left\lvert \sum\limits_{m=1}^{M}\frac{{\rm{e}}^{-{\rm{j}}k_0(\sqrt{(u_k^x-{\psi}_{m})^2+d_k}+n_{\rm{eff}}({\psi}_{m}-{\psi}_{0}^{m})) }}{\sqrt{(u_k^x-{\psi}_{m})^2+d_k}}\right\rvert^2,
\end{align}
\end{subequations}
where ${\bm\psi}\triangleq[{\psi}_{1},\ldots,{\psi}_{M}]^{\mathsf{T}}\in{\mathbbmss{R}}^{M\times1}$ and
\begin{align}
{\mathcal{X}}\triangleq\left\{{\bm\psi}\left\lvert\begin{matrix}{\psi}_{m}\in[\psi_{0}^{m},\psi_{0}^{m}+L],m\in{\mathcal{M}},\\
\lvert{\psi}_{m}-{\psi}_{m'}\rvert\geq\Delta,m\ne m'\end{matrix}\right.\right\}.
\end{align}
Maximizing the per-user rate under SA requires PA locations that produce constructive combining across the segments while maintaining low free-space path loss \cite{xu2024rate,ouyang2025array}. We begin by identifying the segment that contains the projection of the user onto the waveguide. Its index is $m^{\star}=\lceil({u_k^x-{\psi}_{0}^{1}})/{L}\rceil$. The PA in this segment is placed directly beneath the user's projection, i.e., ${\psi}_{m^{\star}}=u_k^x$. Consider next the $(m^{\star}-1)$th segment. To minimize path loss while satisfying the spacing constraint $\Delta$, its initial PA location is chosen as follows:
\begin{align}\label{SA_Optimization_Update_First2}
{\psi}_{m^{\star}-1}=\min\{{\psi}_{0}^{m^{\star}-1}+L,{\psi}_{m^{\star}}-\Delta\}\triangleq {\hat{\psi}}_{m^{\star}-1}.
\end{align} 
We then refine this position to enforce phase alignment with the PA in segment $m^{\star}$ to achieve constructive superposition. The PA at ${\hat{\psi}}_{m^{\star}-1}$ is shifted leftward by $\nu_{m^{\star}-1}>0$ such that
\begin{equation}\nonumber
\begin{split}
&({({\hat{\psi}}_{m^{\star}-1}-\nu_{m^{\star}-1}-u_k^x)^2+d_k})^{1/2}+n_{\rm{eff}}({\hat{\psi}}_{m^{\star}-1}-\nu_{m^{\star}-1}\\
&-{\psi}_{0}^{m^{\star}-1})
=c_{m^{\star}-1}+((c_{m^{\star}}-c_{m^{\star}-1}))\bmod \lambda)\triangleq {\hat{c}}_{m^{\star}-1},
\end{split}
\end{equation}
where $c_{m^{\star}-1}=({(\hat{\psi}_{m^{\star}-1}-u_k^x)^2+d_k})^{1/2}+n_{\rm{eff}}(\hat{\psi}_{m^{\star}-1}-{\psi}_{0}^{m^{\star}-1})$. The closed-form solution for the phase-alignment shift $\nu_{m^{\star}-1}$ is obtained as follows:
\begin{equation}\nonumber
\begin{split}
&\nu_{m^{\star}-1}=\\&\left\{\begin{array}{ll}
\hat{\psi}_{m^{\star}-1}-\frac{{\psi}_{0}^{m^{\star}-1}n_{\rm{eff}}^2+\hat{d}_{m^{\star}-1}n_{\rm{eff}}-u_k^x-\sqrt{\Delta_{m^{\star}-1}}}{n_{\rm{eff}}^2-1}             & {n_{\rm{eff}}\ne1}\\
\hat{\psi}_{m^{\star}-1}-\frac{({\psi}_{0}^{m^{\star}-1}+\hat{d}_{m^{\star}-1})^2-((u_k^x)^2+d_k)}{2({\psi}_{0}^{m^{\star}-1}+\hat{d}_{m^{\star}-1}-u_k^x)}           & {n_{\rm{eff}}=1}
\end{array}\right.,
\end{split}
\end{equation}
where $\Delta_{m^{\star}-1}= n_{\rm{eff}}^2(u_k^x-{\psi}_{m^{\star}-1})^2-2\hat{d}_{m^{\star}-1}n_{\rm{eff}}(u_k^x-{\psi}_{0}^{m^{\star}-1})+d_k(n_{\rm{eff}}^2-1)+\hat{d}_{m^{\star}-1}^2$. Once $\nu_{m^{\star}-1}$ is obtained, the PA position is updated as ${\psi}_{m^{\star}-1}=\hat{\psi}_{m^{\star}-1}-\nu_{m^{\star}-1}$. The PA in the next segment is initialized as follows:
\begin{align}
{\psi}_{m^{\star}-2}=\min\{{\psi}_{0}^{m^{\star}-2}+L,{\psi}_{m^{\star}-1}-\Delta\}\triangleq\hat{\psi}_{m^{\star}-2}, 
\end{align}
and its phase alignment shift is computed using the same procedure as $\nu_{m^{\star}-1}$. This iterative method continues for all segments with $m< m^{\star}$ and extends symmetrically to segments with $m> m^{\star}$ \cite{xu2024rate,ouyang2025array}.
\subsubsection{PM-TDMA}
We now consider PM-based TDMA under the SA protocol. The goal is to maximize the overall sum-rate, which leads to the following optimization problem:
\begin{align}\label{Problem_PM_TDMA_SA}
\argmax\nolimits_{{\bm\psi}\in{\mathcal{X}}}
\sum\nolimits_{k=1}^{K}\log_2(1+{P_k\eta g_k({\bm\psi})}/{(M\sigma^2)}),
\end{align}
where $g_k({\bm\psi})\triangleq \left\lvert \sum\limits_{m=1}^{M}\frac{{\rm{e}}^{-{\rm{j}}k_0(\sqrt{(u_k^x-{\psi}_{m})^2+d_k}+n_{\rm{eff}}({\psi}_{m}-{\psi}_{0}^{m})) }}{\sqrt{(u_k^x-{\psi}_{m})^2+d_k}}\right\rvert^2$. The variables $\{\psi_m\}_{m=1}^{M}$ are coupled through the summation in $g_k({\bm\psi})$. To address this difficulty, we adopt an element-wise alternating optimization method. Each $\psi_m$ is updated sequentially while the remaining variables remain fixed. The subproblem for optimizing $\psi_m$ is formulated as follows:
\begin{subequations}
\begin{align}
\max_{\psi_m}~&f_k^{(m)}(\psi_m)\triangleq\sum_{k=1}^{K}\log_2\left(1+\frac{P_k\eta \lvert\hat{g}_k(\psi_m)+\hat{g}_k^m\rvert^2}{M\sigma^2}\right)\nonumber\\{\rm{s.t.}}~&{\psi}_{m}\in[\psi_{0}^{m},\psi_{0}^{m}+L],\lvert{\psi}_{m}-{\psi}_{m'}\rvert\geq\Delta,m\ne m'\nonumber,
\end{align}
\end{subequations}
where $\hat{g}_k(\psi_m)\triangleq\frac{{\rm{e}}^{-{\rm{j}}k_0(({(u_k^x-{\psi}_{m})^2+d_k})^{1/2}+n_{\rm{eff}}({\psi}_{m}-{\psi}_{0}^{m})) }}{({(u_k^x-{\psi}_{m})^2+d_k})^{1/2}}$ and $\hat{g}_k^m\triangleq\sum_{m'\ne m}\frac{{\rm{e}}^{-{\rm{j}}k_0(({(u_k^x-{\psi}_{m'})^2+d_k})^{1/2}+n_{\rm{eff}}({\psi}_{m'}-{\psi}_{0}^{m'})) }}{({(u_k^x-{\psi}_{m'})^2+d_k})^{1/2}}$. Since the subproblem contains a single variable over a bounded interval, it can be solved efficiently through \emph{one-dimensional search}. To implement the search, the interval $[\psi_{0}^{m},\psi_{0}^{m}+L]$ is discretized into the following uniform grid of $Q$ points:
\begin{align}\nonumber
{\mathcal{Q}}_m\triangleq \left\{\psi_{0}^{m},\psi_{0}^{m}+\frac{L}{Q-1},\psi_{0}^{m}+\frac{2L}{Q-1},\ldots,\psi_{0}^{m}+L\right\}.
\end{align}
Because of the minimum inter-antenna spacing constraint, some grid points become infeasible. We therefore define
\begin{align}
\hat{\mathcal{Q}}_m\triangleq \{x|x\in{\mathcal{Q}}_m,\lvert x - \psi_{m'}\rvert<\Delta,m'\ne m\},
\end{align}
and select a near-optimal $\psi_m$ as follows:
\begin{align}\label{MP_Pareto_Sub_sub1}
\argmax\nolimits_{\psi_m\in {\mathcal{Q}}_m\setminus \hat{\mathcal{Q}}_m}{f_k^{(m)}(\psi_m)}.
\end{align}
This update is applied sequentially to all PAs until convergence. The complete algorithm is summarized in Algorithm \ref{Algorithm1}, with a computational complexity of ${\mathcal{O}}(I_{\rm{iter}} MKQ)$, where $I_{\rm{iter}}$ is the number of iterations to convergence.

\begin{algorithm}[!t]  
\algsetup{linenosize=\tiny} \scriptsize
\caption{Element-wise Algorithm for Solving \eqref{Problem_PM_TDMA_SA}}
\label{Algorithm1}
\begin{algorithmic}[1]
\STATE initialize the optimization variables
\REPEAT 
  \FOR{$m\in\{1,\ldots,M\}$}
      \STATE update $\psi_m$ through one-dimensional search
    \ENDFOR
\UNTIL{the fractional increase of the objective value falls below a predefined threshold}
\end{algorithmic}
\end{algorithm}
\subsubsection{NOMA}
We now examine the case of NOMA-based SA. The objective is to maximize the sum-rate ${\mathcal{R}}_{\rm{N}}$, which yields
\begin{align}\label{Problem_NOMA_SA}
\max_{{\bm\psi}\in{\mathcal{X}}}
\sum_{k=1}^{K}P_k \left\lvert \sum\limits_{m=1}^{M}\frac{{\rm{e}}^{-{\rm{j}}k_0(\sqrt{(u_k^x-{\psi}_{m})^2+d_k}+n_{\rm{eff}}({\psi}_{m}-{\psi}_{0}^{m})) }}{\sqrt{(u_k^x-{\psi}_{m})^2+d_k}}\right\rvert^2.
\end{align}
The variables $\{\psi_m\}_{m=1}^{M}$ remain coupled through the summation. An element-wise alternating optimization method is therefore suitable. Each $\psi_m$ is updated sequentially while all other PA locations remain fixed. The subproblem for optimizing $\psi_m$ can be formulated as follows:
\begin{subequations}
\begin{align}
\max_{\psi_m}~&{\hat{f}}_k^{(m)}(\psi_m)\triangleq\sum\nolimits_{k=1}^{K}P_k\lvert\hat{g}_k(\psi_m)+\hat{g}_k^m\rvert^2\nonumber\\{\rm{s.t.}}~&{\psi}_{m}\in[\psi_{0}^{m},\psi_{0}^{m}+L],\lvert{\psi}_{m}-{\psi}_{m'}\rvert\geq\Delta,m\ne m'\nonumber,
\end{align}
\end{subequations}
whose optimal solution can be approximated via the following grid sarch:
\begin{align}\label{MP_Pareto_Sub_sub1}
\argmax\nolimits_{\psi_m\in {\mathcal{Q}}_m\setminus \hat{\mathcal{Q}}_m}{{\hat{f}}_k^{(m)}(\psi_m)}.
\end{align}
The updates continue for all antennas until convergence. The complete procedure is similar to Algorithm \ref{Algorithm1} and has a computational complexity of ${\mathcal{O}}(I_{\rm{iter}} MKQ)$. For reference, Table \ref{Table: PASS_Uplink} summarizes the protocols considered for uplink multiuser SWAN together with the proposed optimization approaches.

\begin{table*}[!t]
\centering
\caption{Summary of the Proposed Protocols for Uplink SWAN Multiuser Communications}
\setlength{\abovecaptionskip}{0pt}
\resizebox{0.95\textwidth}{!}{
\begin{tabular}{|c|cc|c|}
\hline
\multirow{2}{*}{\textbf{Protocol}} & \multicolumn{2}{c|}{\textbf{TDMA (Time Division Multiple Access)}}                                               & \multirow{2}{*}{\textbf{NOMA (Non-Orthogonal Multiple Access)}} \\ \cline{2-3}
                          & \multicolumn{1}{c|}{\textbf{PS (Pinch Switching)}}                                            & \textbf{PM (Pinch Multiplexing)} &                       \\ \hline
\textbf{SS (Segment Selection)}                        & \multicolumn{1}{c|}{Problem \eqref{PS_TDMA_SS} (\emph{closed-form}, optimal)}     & Problem \eqref{PM_TDMA_SS} (\emph{one-dimensional search}, near-optimal)  & Problem \eqref{NOMA_SS} (\emph{one-dimensional search}, near-optimal)                \\ \hline
\textbf{SA (Segment Aggregation)}                        & \multicolumn{1}{c|}{Problem \eqref{PS_TDMA_SA} (\emph{antenna position refinement}, optimal)} & Problem \eqref{Problem_PM_TDMA_SA} (\emph{element-wise alternating optimization}, sub-optimal)  & Problem \eqref{Problem_NOMA_SA} (\emph{element-wise alternating optimization}, sub-optimal)              \\ \hline
\end{tabular}}
\label{Table: PASS_Uplink}
\vspace{-10pt}
\end{table*}

\section{Numerical Results}\label{Section: Numerical Results}
We provide numerical results to validate the effectiveness of the proposed algorithms and to compare the performance of the considered protocols. Unless stated otherwise, the simulation parameters are as follows: carrier frequency $f_{\rm{c}} = 28$ GHz, effective refractive index $n_{\rm{eff}}=1.4$, minimum inter-antenna spacing $\Delta=\frac{\lambda}{2}$, waveguide deployment height $d=3$ m, segment length $L=1$ m, transmit power $P_k=10$ dBm for $k\in{\mathcal{K}}$, noise power $\sigma^2=-90$ dBm, and grid search precision $Q=10^4$. There are $K=4$ users uniformly distributed in a rectangular region centered at the origin with side lengths $D_x=50$ m and $D_y = 20$ m. The $x$-coordinate of the feed point of the first waveguide segment is $\psi_{0}^{1}=-\frac{D_x}{2}$. For comparison, the performance of SWAN is evaluated against a conventional single-waveguide PASS architecture where only one PA is activated along the waveguide. The case of conventional PASS with multiple active PAs is excluded because no tractable uplink model exists due to the IAR effect. For the proposed element-wise alternating optimization methods, the PAs are initialized at the center of each segment.

\begin{figure}[!t]
\centering
    \subfigure[Case \uppercase\expandafter{\romannumeral1}.]
    {
        \includegraphics[height=0.17\textwidth]{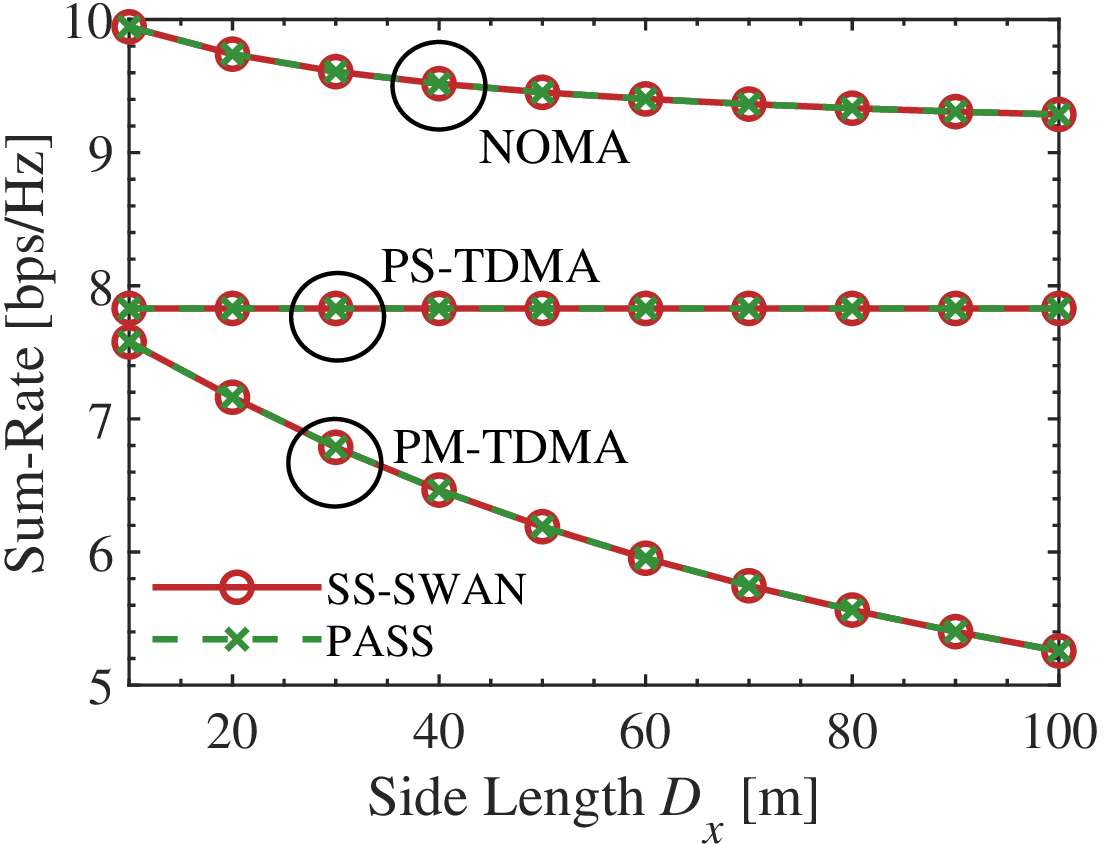}
	   \label{fig:placeholder1}
    }
    \subfigure[Case \uppercase\expandafter{\romannumeral2}.]
    {
        \includegraphics[height=0.17\textwidth]{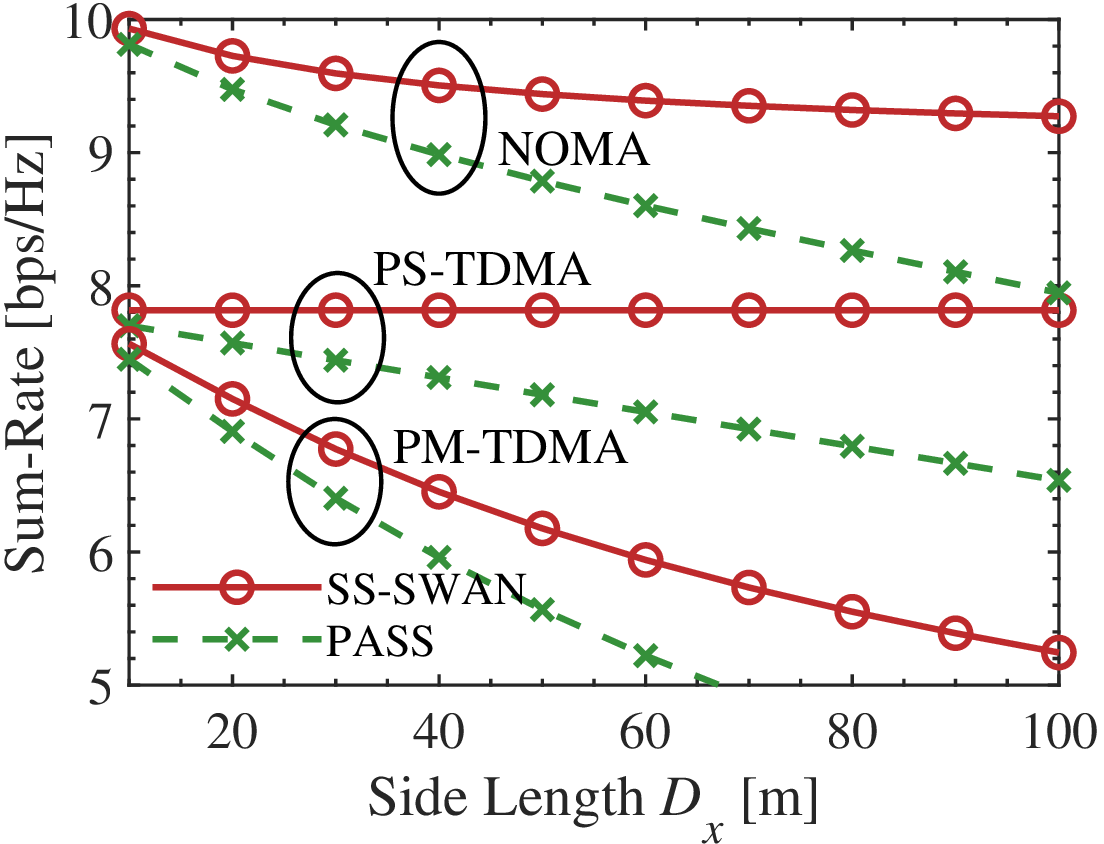}
	   \label{fig:placeholder2}
    }
\caption{Comparison of SWAN and conventional PASS.}
\label{fig:placeholder}
\vspace{-15pt}
\end{figure}

{\figurename} {\ref{fig:placeholder}} compares the sum-rates achieved by SS-based SWAN and conventional PASS under both TDMA and NOMA. For completeness, two cases are included: \romannumeral1) no in-waveguide propagation loss with $\kappa=0$ (Case \uppercase\expandafter{\romannumeral1} in {\figurename} {\ref{fig:placeholder1}}), and \romannumeral2) in-waveguide loss with an average attenuation factor of $\kappa=0.08$ dB/m (Case \uppercase\expandafter{\romannumeral2} in {\figurename} {\ref{fig:placeholder2}}) \cite{ding2024flexible}. As shown, when in-waveguide loss is ignored, conventional PASS achieves the same performance as SWAN. This result is expected because both architectures activate only a single PA, and the optimal PA placements should be the same. Once in-waveguide loss is taken into account, however, conventional PASS yields a noticeably lower sum-rate than SS-based SWAN, and the performance gap becomes more pronounced as the side length $D_x$. This behavior is reasonable. In conventional PASS, the average distance between the users and the feed point increases monotonically with $D_x$, which leads to stronger in-waveguide attenuation and a substantial reduction in the achievable sum-rate. In contrast, SWAN confines in-waveguide propagation to the segment level rather than the entire waveguide. Each segment possesses an independent feed point that connects directly to the BS. As a result, SWAN maintains relatively stable performance and experiences negligible degradation from in-waveguide loss even when the side length grows, which can be seen by comparing {\figurename} {\ref{fig:placeholder1}} with {\figurename} {\ref{fig:placeholder2}}.

{\figurename} {\ref{fig:placeholder}} also shows that NOMA achieves the highest sum-rate, which is as expected. Moreover, PS-TDMA outperforms PM-TDMA because PS allows the PA to be positioned optimally for each user within its time slot. Another observation is that the sum-rate achieved by PS-TDMA remains nearly constant with respect to $D_x$. This arises because PS optimizes the PA location for each user individually, which enables the PA to be placed close to the corresponding user regardless of the user's position in the service region. The results highlight the performance advantage of SWAN in reducing both large-scale path loss and in-waveguide attenuation.

\begin{figure}[!t]
\centering
    \subfigure[$D_x=100$ m.]
    {
        \includegraphics[height=0.17\textwidth]{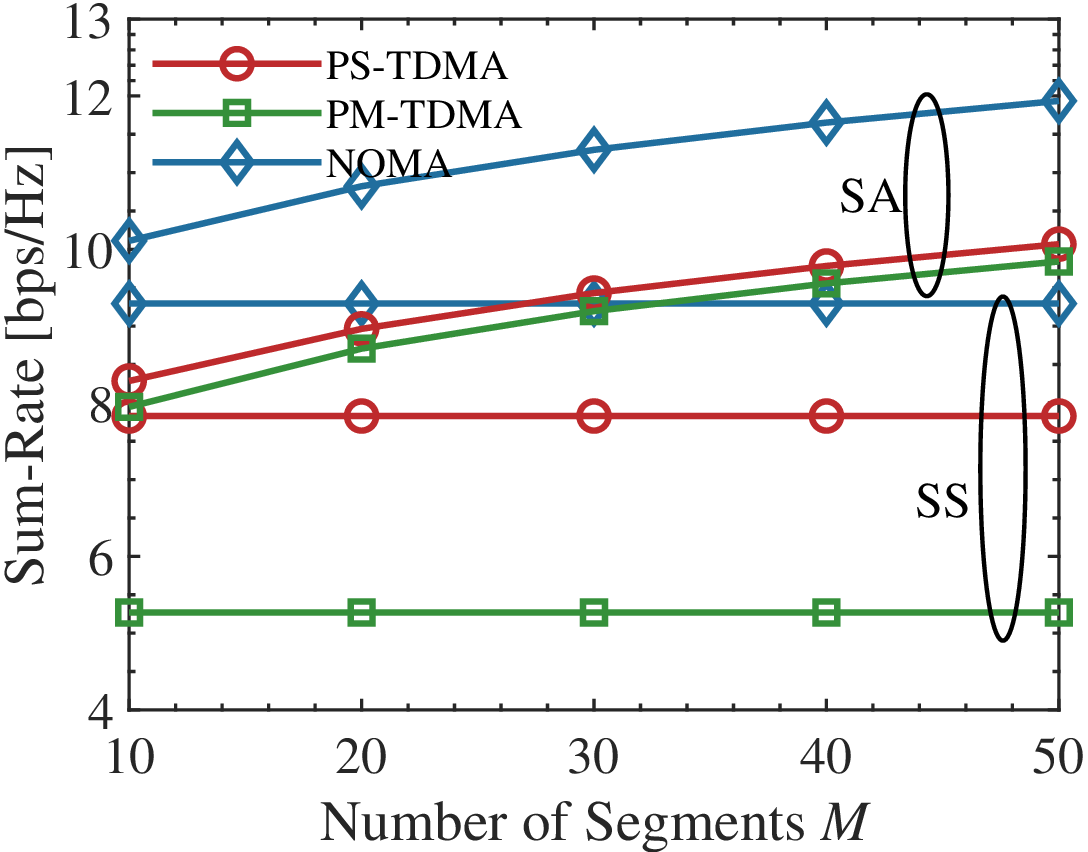}
	   \label{fig:waveguide_performance}
    }
    \subfigure[$L=1$ m.]
    {
        \includegraphics[height=0.17\textwidth]{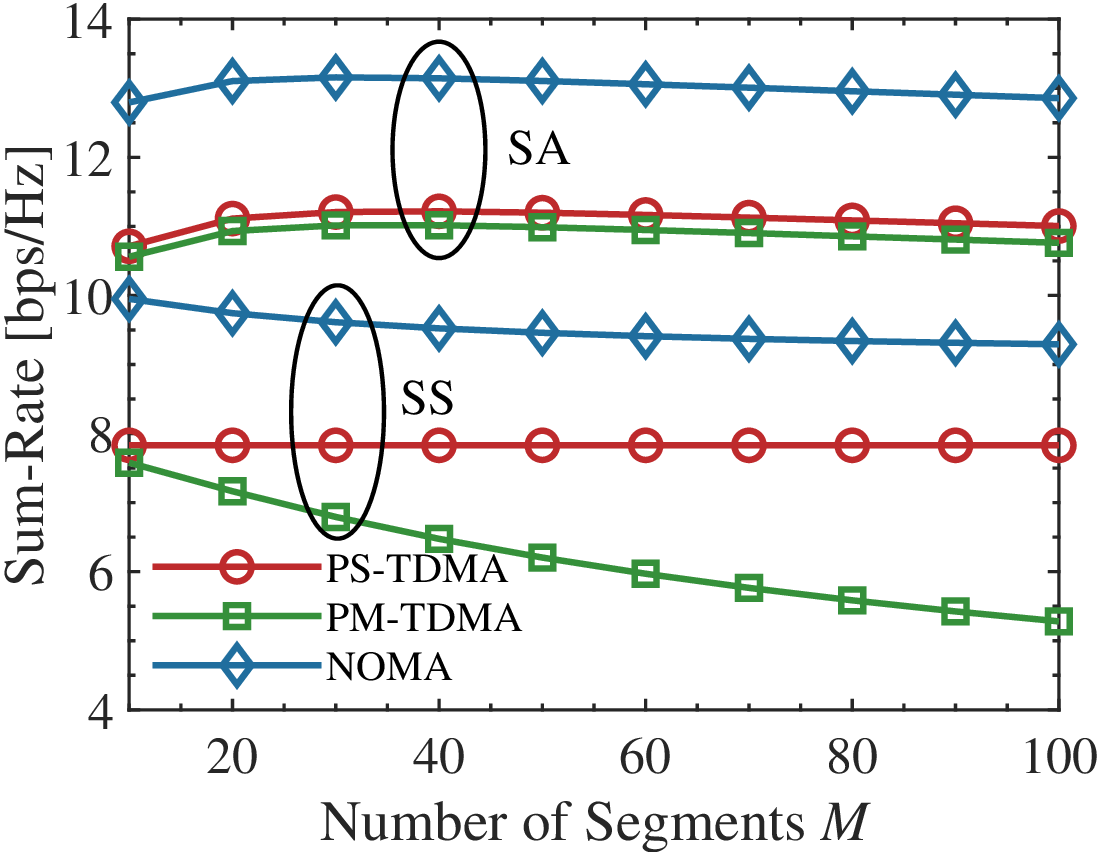}
	   \label{fig:fixed_segment_results}
    }
\caption{Comparison of SS and SA.}
\vspace{-15pt}
\end{figure}

Since the results above have demonstrated that SWAN outperforms conventional PASS and that in-waveguide propagation loss has a negligible impact on the performance of SWAN, the following figures focus solely on SWAN and ignore in-waveguide attenuation. {\figurename} {\ref{fig:waveguide_performance}} compares the performance of SS and SA with respect to the number of segments while the side length is fixed at $D_x=100$ m. The figure shows that the performance of SA improves as the number of segments increases. This improvement occurs because more segments provide more PAs, which results in a larger effective array. The figure also shows that SA outperforms SS under the same multiple-access scheme (TDMA or NOMA) because SA captures additional array gain. A noteworthy observation is that SS-based NOMA can even surpass SA-based TDMA in terms of achievable sum-rate. This highlights the strength of NOMA in managing inter-user interference and improving spectral efficiency. {\figurename} {\ref{fig:fixed_segment_results}} further illustrates the sum-rate as a function of the number of segments when the segment length is fixed at $L=1$ m. The sum-rate achieved by SA increases at first as the number of segments grows, reaches a maximum, and then decreases. This trend is consistent with the theoretical results in \cite{ouyang2025array}, which suggest that an optimal number of segments exists when the segment length is fixed. Hence, both the number of segments and the PA positions should be jointly optimized to maximize the sum-rate.

\begin{figure}
    \centering
    \includegraphics[height=0.17\textwidth]{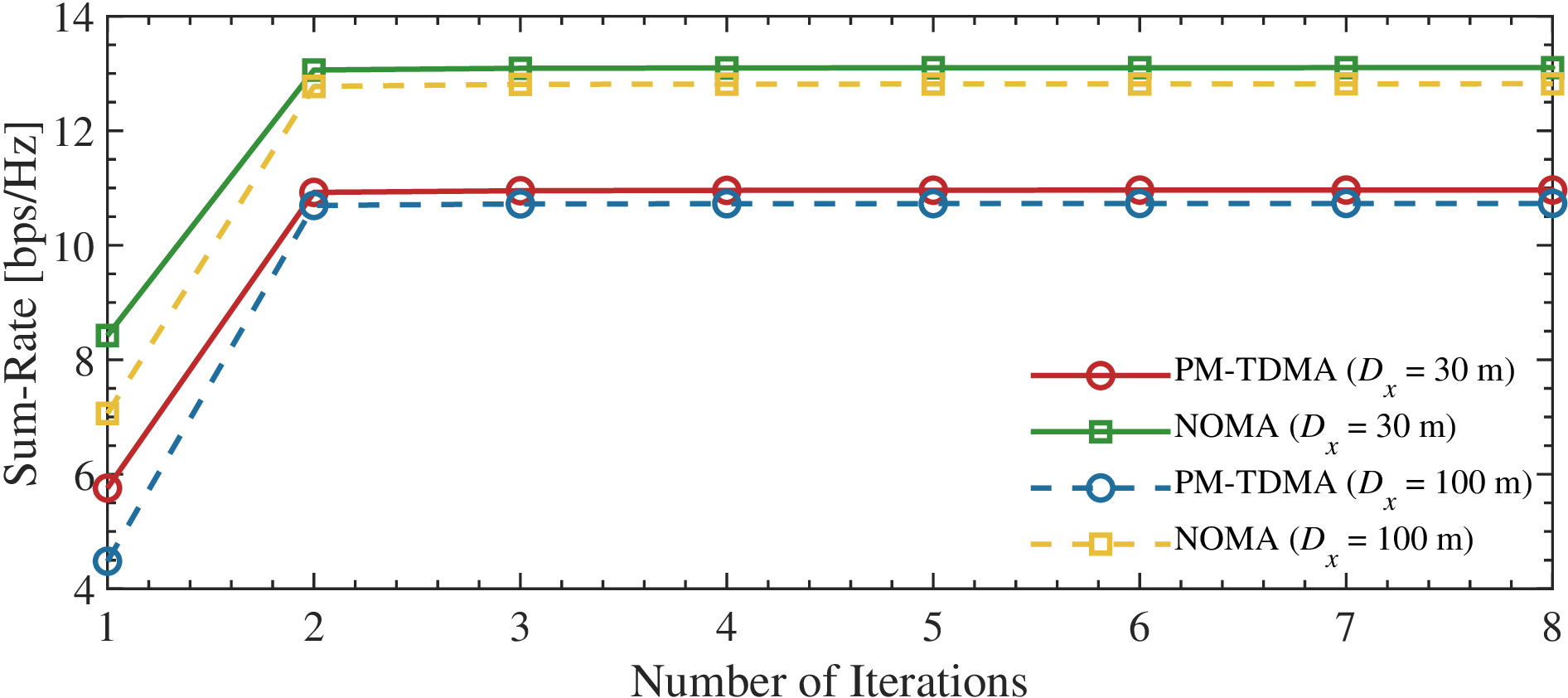}
    \caption{Convergence of the proposed methods.}
    \label{fig:pa_performance}
    \vspace{-10pt}
\end{figure}

Finally, {\figurename} {\ref{fig:pa_performance}} illustrates the convergence behavior of the proposed element-wise alternating optimization method for PA placement under SA. This figure shows that the algorithm converges in fewer than five iterations, which indicates low computational complexity. Combined with the previous results that demonstrate strong sum-rate performance for both the TDMA and NOMA cases, we conclude that the proposed element-wise alternating optimization framework is an effective and practical solution for multiuser SWAN communications.

\section{Conclusion}
A SWAN-based uplink multiuser transmission framework has been proposed. The achievable sum-rate was maximized by optimizing the PA placement under both SS and SA protocols for TDMA and NOMA. The results showed that SWAN outperforms conventional uplink PASS in terms of sum-rate. The results also revealed the existence of an optimal number of segments that maximizes the sum-rate, which suggests a promising future research direction on the joint optimization of the number of segments and the PA placement.
\bibliographystyle{IEEEtran}
\bibliography{mybib}
\end{document}